\documentclass{birkjour}

 \theoremstyle{definition}
 
 \theoremstyle{remark}

 \numberwithin{equation}{section}
\usepackage{url}
\usepackage{amsmath,amssymb}
\usepackage{amsfonts}
\usepackage{subfig} 
\usepackage{caption}

\begin{document}

\title[Impact of contamination factors on the COVID-19 evolution in Senegal]
 {Impact of contamination factors on the COVID-19 evolution in Senegal}
 
\author[Ndiaye V.M., Sarr O.S.]{Vieux Medoune Ndiaye, Serigne Omar Sarr}
\address{Faculté de Médecine, de Pharmacie et d'Odontologie, \br
University of Cheikh Anta Diop, Dakar, Senegal}
\email{vieuxmedoune.ndiaye@ucad.edu.sn, serigne.sarr@ucad.edu.sn}

\author[Ndiaye B.M.]{Babacar Mbaye Ndiaye}
\address{Laboratory of Mathematics of Decision and Numerical Analysis\br
University of Cheikh Anta Diop\br
BP 45087, 10700. Dakar, Senegal}
\email{babacarm.ndiaye@ucad.edu.sn}


\date{June 17, 2020}
\begin{abstract}
In this article, we perform an analysis of COVID-19 on one of the South Saharan countries (hot zone), the Senegal (West Africa). Many questions remain unanswered: why the African continent is not very contaminated compared to other continents.
Factors of cross immunity, temperature, population density, youth, etc. are taken into account for an analysis of the contamination factors.
Numerical simulations are carried out for a prediction over the coming week. 
\end{abstract}

\maketitle

\section{Introduction}
\noindent  The COVID-19 pandemic is the new planetary threat at the start of 2020. Since its appearance in December 2019 in Wuhan in Hubei province in China, the pandemic now affects all five continents.
The impressive multiplication of confirmed cases between Europe and America has caused an oscillation of the focus between these two continents notwithstanding health systems (but renowned for their performance).\\
At the beginning of the epidemic, health authorities and international institutions believe that the pandemic would inevitably be devastating for the African continent due to the fragile health systems and insufficient medical equipment.\\
If certain African countries have to face many cases of contamination, like in Egypt and South Africa, for a large part of the other countries the wave is less important than expected with a globally different kinetics.\\
In Senegal, the threshold of 4851 cases was crossed on June 12th, 2020 with 1839 infected and 56 deaths. This assessment is certainly alarming but less important compared to certain countries of Europe and America and far from gravitating around the forecasts of the WHO and certain experts. Obviously the low number of tests and the lack of data partially skew the balance sheet but behind these figures, there are probably factors influencing the slow speed of spread of the virus.\\
These factors are worth considering. It is in this context that this work is fixed as a goal of reflection on the various and varied factors that can explain this particular kinetics of the epidemic in Senegal.\\
\noindent First, we collect the pandemic data from \cite{msas},  
  from March 03, 2020 to June 12, 2020. In the second step, we propose some contamination factors and machine learning technics for a week forecasting.\\
\noindent The article is organized as follows. In section \ref{analysis}, we present some data analysis followed by contamination factors in section \ref{factors}. In section \ref{simnum},  we perform machine learning technics for forecasting for the following week.  Finally, in the section \ref{ccl}, we present conclusions and perspectives.

\section{Data analysis}\label{analysis}
\begin{figure}[h!]
  \subfloat[Senegal cases - confirmed, recovered and deaths]{
	\begin{minipage}[1\width]{0.5\textwidth}
	   \centering
	   \includegraphics[width=1.25\textwidth]{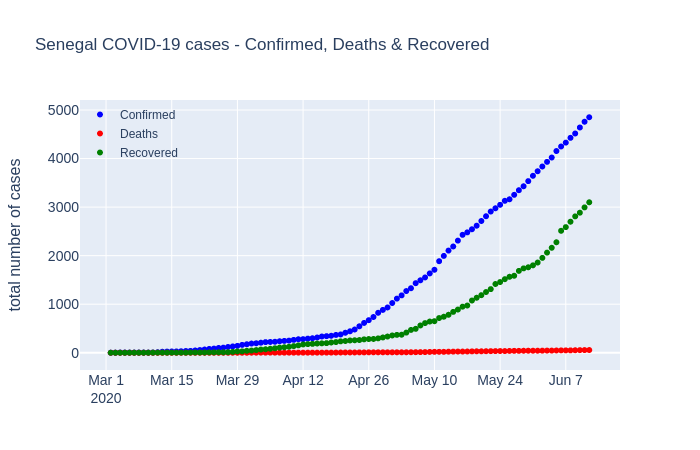}\label{crd_sn}
	\end{minipage}}
  \subfloat[Dakar zones \cite{ausn}]{
	\begin{minipage}[1\width]{ 0.5\textwidth}
	   \centering
	   \includegraphics[width=.9\textwidth]{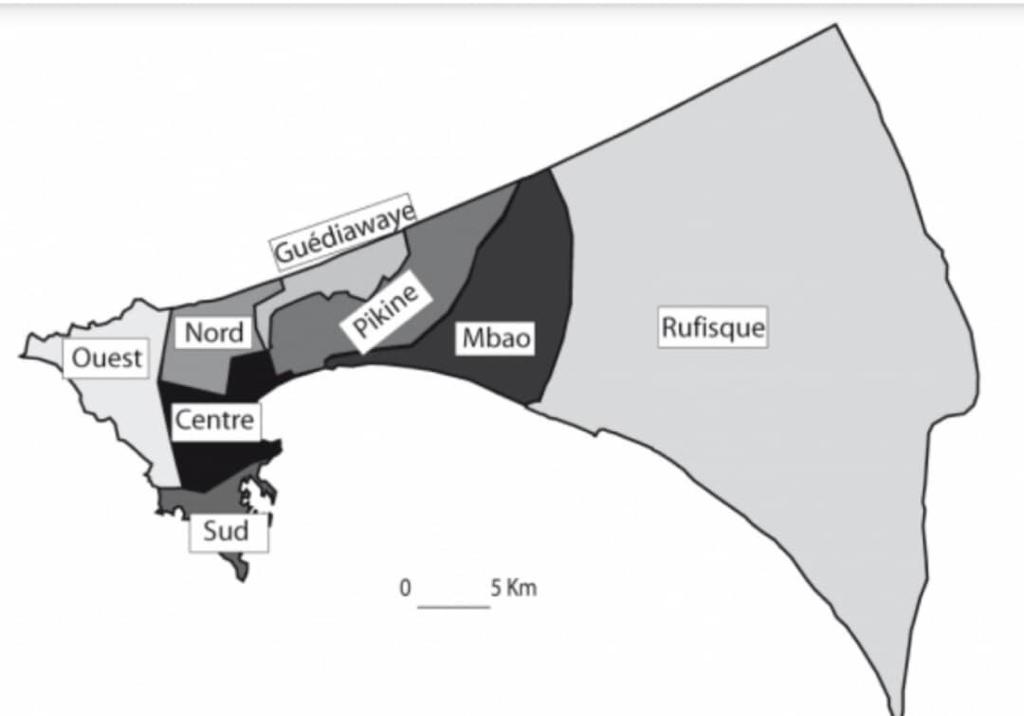}\label{dkrZones}
	\end{minipage}}
\caption{Senegal: community and severe cases}\label{testsresults}
\end{figure}

The simulations are carried out from  data in \cite{msas}, from March 02, 2020 to June 12, 2020. 
The numerical tests were performed by using the Python with Panda library 
\cite{python}, and were executed on a computer with following characteristics: intel(R) Core-i7 CPU 2.60GHz, 24.0Gb of RAM, under UNIX system.\\
According to daily reports, we first analyze and make some data preprocessing before simulations. The cumulative numbers of confirmed, recovered and deaths cases are illustrated in Figure \ref{crd_sn} and Figure \ref{dkrZones} illustrates Dakar zones (with higher confirmed cases (see section \ref{popdensity})). \noindent In Figure \ref{dkrZones}, "Dakar Ouest" = East of Dakar, "Dakar Nord" = North of Dakar, "Dakar centre" = Dakar center and "Dakar sud"  = South of Dakar.
 \\
We get various summary statistics (per day), by giving the mean, standard deviation, minimum and maximum values, and the quantiles of the data (see Tables \ref{stat_confcases_1} and \ref{stat_confcases_2}). 
\begin{table}[h!] 
\begin{center}
\begin{tabular}{|c|c|c|c|c|c|c|} 
 \hline
 {\bf values} 	& {\bf tests} & 	{\bf cases} & {\bf	contact} &{\bf	imported}& 	{\bf community}& 	{\bf confirmed} \\ \hline
mean 	& 577.44 	&47.09  &	41.06  &	0.96 &	5.07 &	1332.50\\ 
\hline
std &	507.77 &	43.48 & 	39.97  &	1.92	&5.93  &	1512.43  \\ 
\hline
min 	&1 &	0 	&0& 	0 &	0 &	1  \\ 
\hline
25\% 	&97.25 &	7.5 &	3.5 &	0 &	0 &	124.5 \\ 
\hline
50\% &	474 &	31& 	28& 	0& 	3 &	442 \\ 
\hline
75\% &	1008.75& 	89 &	75.5 &	1& 	8 &	2512  \\ 
\hline
max &	1820 &	177& 	169 &	11 &	30 &	4851 \\
\hline
\end{tabular}
\end{center}
\caption{Senegal summary statistics (per day) until June 12th, 2020 (tests, cases, 	contact, imported,  community and confirmed).}\label{stat_confcases_1}
\end{table}
\begin{table}[h!] 
\begin{center}
\begin{tabular}{|c|c|c|c|c|c|c|} 
 \hline
 {\bf values} 	& {\bf recovered} &{\bf deaths} 	& {\bf evacuated} 	& {\bf severes} &	{\bf infected} &	{\bf ratios} \\ \hline
mean 	&650.84 	&14.44 & 	0.69  &	4.93  &	667.22 &	0.14 \\ 
\hline
std &	842.50 &	17.52 &	0.46 &	6.68   &	687.49 &	0.22  \\ 
\hline
min 	&  	0 	& 0 &	0  &	0& 	1 &	0 \\ 
\hline
25\% 	& 	14.5 &	0& 	0& 	0 &	107.5 &	0.062426 \\ 
\hline
50\% &253 &	6 &	1& 	0 &	183 &	0.080638 \\ 
\hline
75\% &	1024.5 &	27 	&1& 	9 &	1454.5& 	0.102737 \\ 
\hline
max & 3100 &	56 &	1 &	25& 	1839  &	1 \\
\hline
\end{tabular}
\end{center}
\caption{Senegal summary statistics (per day) until June 12th, 2020 (recovered, deaths, evacuated, severe, infected and ratios).}\label{stat_confcases_2}
\end{table}

\begin{figure}[h!]
  \subfloat[number of tests]{
	\begin{minipage}[1\width]{0.5\textwidth}
	   \centering
	   \includegraphics[width=1.1\textwidth]{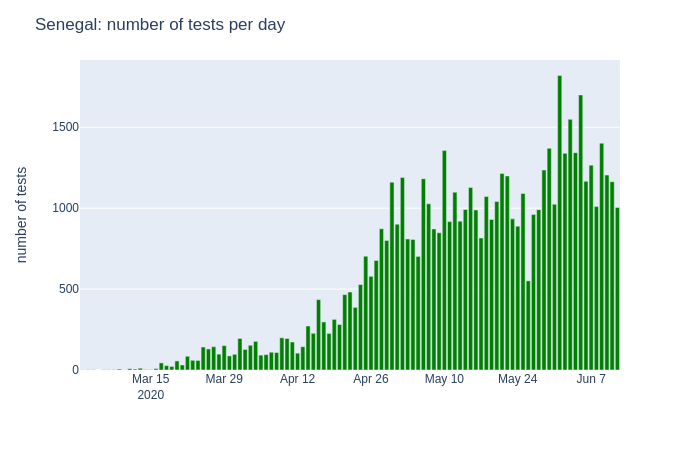}\label{tests}
	\end{minipage}}
  \subfloat[ratios]{
	\begin{minipage}[1\width]{ 0.5\textwidth}
	   \centering
	   \includegraphics[width=1.1\textwidth]{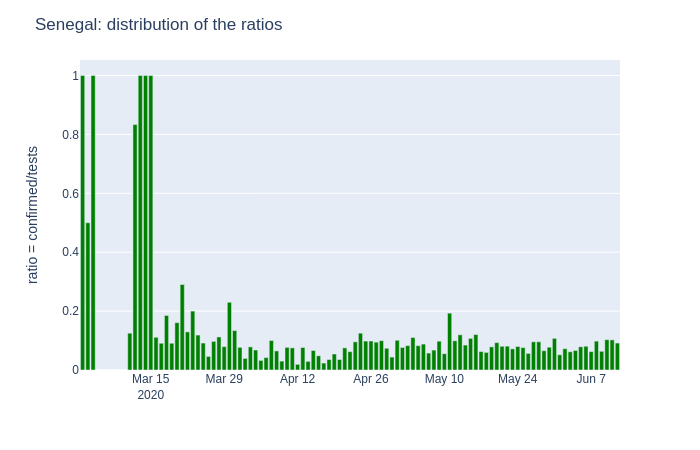}\label{ratios}
	\end{minipage}}
\newline
  \subfloat[severe cases]{
	\begin{minipage}[1\width]{ 0.5\textwidth}
	   \centering
	   \includegraphics[width=1.1\textwidth]{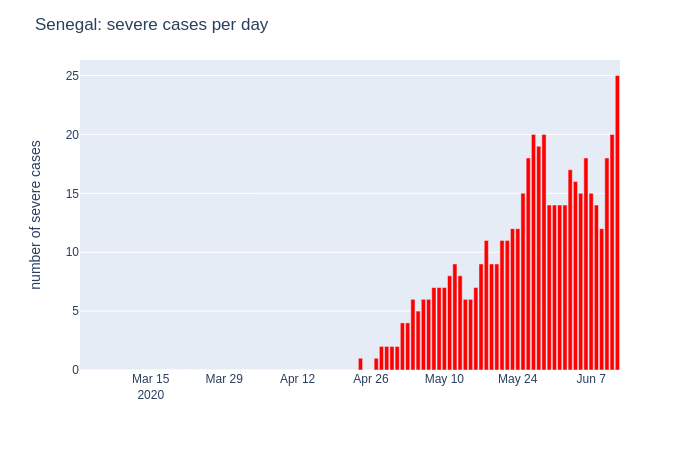}\label{severe}
	\end{minipage}}
%
  \subfloat[community cases]{
	\begin{minipage}[1\width]{ 0.5\textwidth}
	   \centering
	   \includegraphics[width=1.1\textwidth]{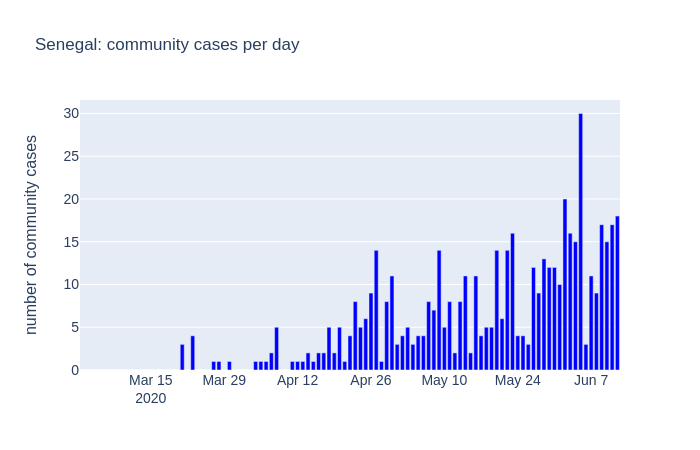}\label{comty}
	\end{minipage}}
\caption{Senegal: community and severe cases}\label{testsresults}
\end{figure}

\noindent The number of performed tests par day is not a constant (see Figure \ref{tests}). 
The Figure \ref{ratios} shows that in Senegal, between March 02 and June 12, 2020, the ratio  (confirmed cases/tests) is almost constant despite variations in the number of daily tests performed and considered to be unrepresentative. Despite official statements about the peak period, it seems difficult to say whether the peak of contamination has already been reached or will even be reached shortly. 
In addition, we find that the ratio varies on average by 10\%. \\
Figures \ref{severe} and \ref{comty} show that the maximum number of severe and community cases are less than 10\% of the number of confirmed cases.

\section{The indicator factors}\label{factors}
In this section, we describe some natural indicator factors which can limit the spread of the virus in Senegal.
\subsection{Temperature - Humidity: guaranteed sun all year round}
Actually, The COVID-19 period coincides with the dry season. 
Senegal is one of the sunniest countries in the world: more than 3,000 hours of sunshine a year.
There are two seasons:
\begin{itemize}
\item a rainy season, from June to October, with greater precipitation from south to north;
\item a dry season, from November to May, with temperatures between 22$^\circ$C and 30$^\circ$C, and significant variations between the coast and the interior.
\end{itemize}
The evolution from North to South: 
\begin{itemize}
\item in Dakar, the average daytime maximum is 24$^\circ$C from January to March and between 25 and 27$^\circ$C in April, May and December. From June to October, temperatures reach 30$^\circ$C.
\item in southern Senegal, the coolest period is from December to mid-February, with daytime averages of around 24$^\circ$C.
\end{itemize}
In October and November, and from mid-February to April, maximum temperatures are around 26$^\circ$C. From July to September, they reach 30$^\circ$C.\\
\noindent The amount of precipitation increases from north to south of the country. In the far north (Senegal river region), the average annual rainfall is 300 mm, while in the far south (lower Casamance, Kolda region), it can exceed 1,500 mm. 

\begin{figure}[h!]
	\centering
	\includegraphics[width=.6\linewidth]{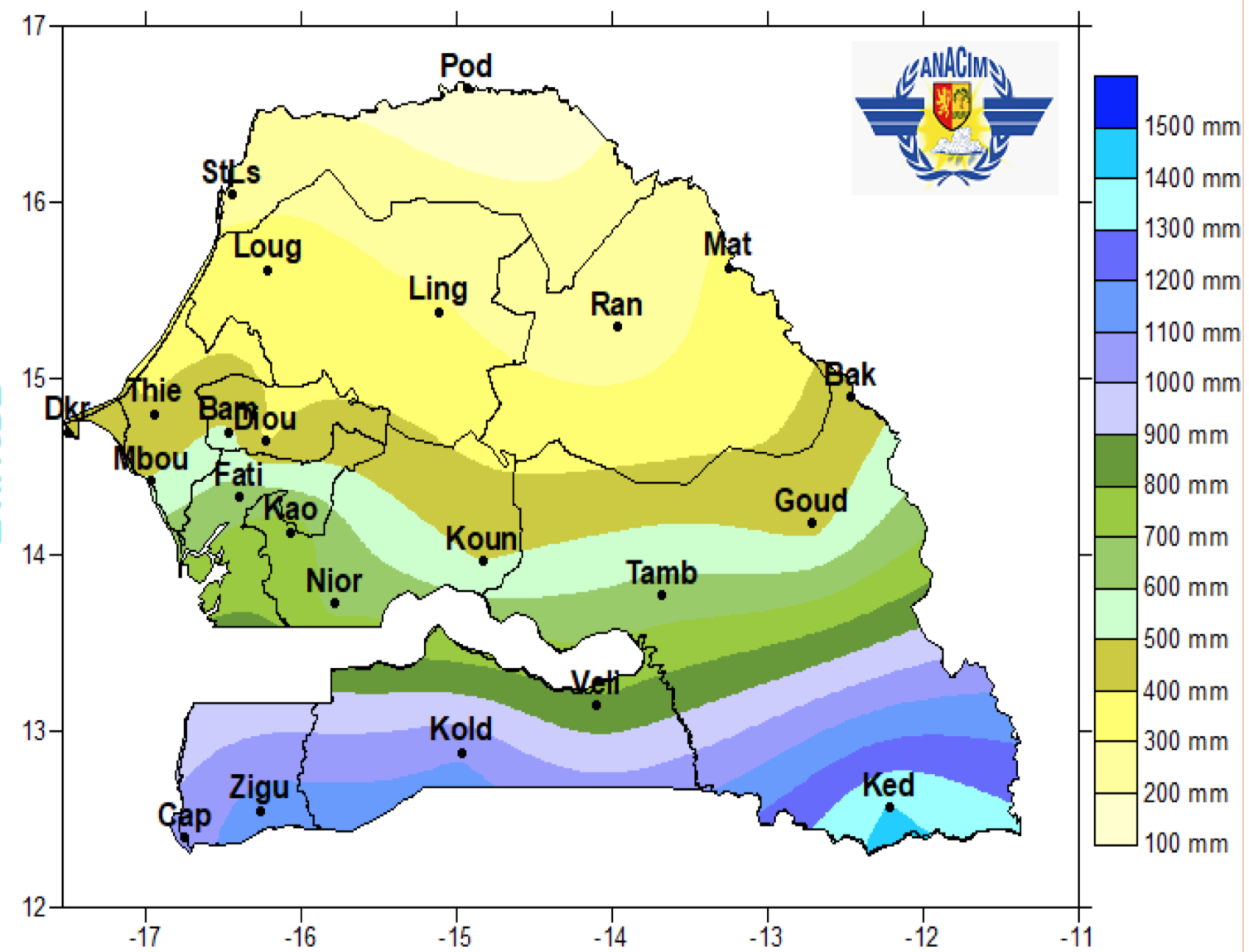}
	\caption{Cumulative rainfall by December 31, 2019 \cite{pluie1}}\label{pluvio}
\end{figure}

\begin{figure}[h!]
  \subfloat[Confirmed cases in all regions]{
	\begin{minipage}[1\width]{0.5\textwidth}
	   \centering
	   \includegraphics[width=1.\textwidth]{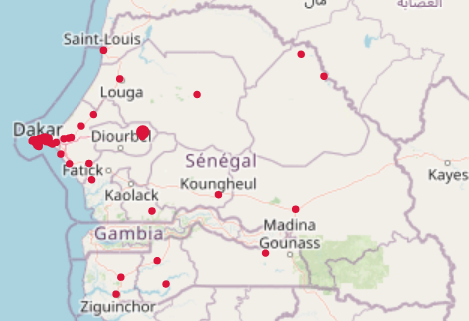}\label{regions_all}
	\end{minipage}}
  \subfloat[Zoom on confirmed cases in regions]{
	\begin{minipage}[1\width]{ 0.5\textwidth}
	   \centering
	   \includegraphics[width=1.\textwidth]{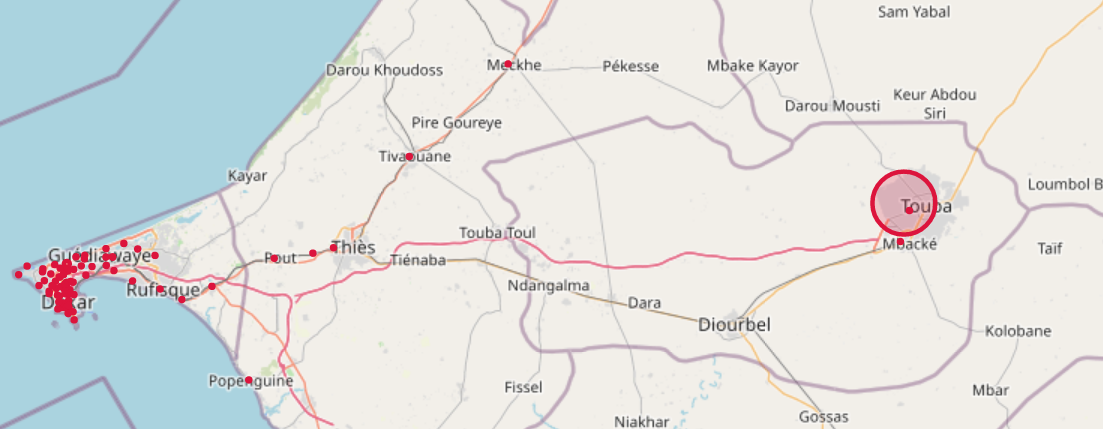}\label{regions_zoom1}
	\end{minipage}}
\caption{Senegal: visualization of confirmed cases from North to South}\label{sn_rainfull}
\end{figure}

\noindent The cumulative rainfall values from North to South are given in Figure \ref{pluvio}, and Figure in \ref{sn_rainfull} shows the mapping of COVID-19 in Senegal as of June 12, 2020.


\noindent The hypothesis seems to be corroborated by the fact that the regions most affected by the pandemic have a rather temperate climate and that most of the cases are concentrated either in the extreme west of the country or in the extreme south. 
This hypothesis is in line with the work in \cite{R-value}, where they predict a sharp decline in the disease from July in Senegal.

\subsection{	Youth}
The population of Senegal is young and varied. We have 54\% of the population who is under the age of 20 \cite{population2}, whose annual growth rate is 3.8\%. 
Life expectancy at birth (year 2017) is 64. 
\noindent Young people play a lot of sport. The relative youth of the Senegalese population with fewer elderly people can really explain the low mortality rate.

\subsection{Cross immunity}
Cross immunity describes acquired immunity against an infectious agent that protects against another agent (virus or bacteria). Cross immunity is linked to the phenomenon of cross reaction. In general, an antibody is specific for an antigen; but sometimes antibodies bind to close antigens (they are called cross-reactive), because these antigens have common epitomes or are of similar structure.
People who have already encountered certain influenza viruses in their life would be better protected than others against other influenza viruses (\cite{immunity1,immunity2,immunity3}).
\noindent In Senegal, the seasonal flu virus is known to mutate each year, but these variations are often minor. This explains why there may be a share of cross immunity with viruses encountered in previous years.\\
Nevertheless, this new type of coronavirus (COVID-19) is caused by a new type of coronavirus first named 2019-nCoV, then renamed SARS-CoV-2, never seen before.\\
\noindent The Senegalese have known certain diseases for a very long time. Malaria occurs throughout the year and across the country. Bilharziasis is especially present in Casamance (south of Senegal), allergies: dust, pollen throughout the year, intestinal parasites are also very common.\\
\noindent However, vaccination campaigns are carried out (certain vaccines are optional) by the Ministry of Health \cite {msas}:
yellow fever, hepatitis A, tetanus, polio, diphtheria, BCG, MMR (Measles, Mumps, Rubella), whooping cough, meningitis A and C, typhoid.

\subsection{Food}
\subsubsection{Meals}
The poverty of many urban and rural families does not always allow them to prepare two meals a day and to diversify their diet. They are often satisfied with a bowl of rice or millet, sometimes with a few vegetables, a little fish, a piece of meat to share among the many members of the family. Senegal is a country where fish is widely consumed. Garlic, bay leaf, etc. are used for the taste of the kitchen.
The Senegalese diet consists mainly of cereals (staple food): rice, wheat, millet and fonio. \\
In addition, some senegalese dishes like "lakhou thiakhane", "gourbane", senegalese soup,  etc. are recommended foods for the flu and malaria.

\subsubsection{Fruits}
\noindent The main fruits are mangoes, which are part of the landscape. There are also grapefruits (roses are better), papayas, oranges (often not very juicy, but good - consume Senegalese - pressed between the teeth), melons, corrosols and guavas, with a delicate and exotic fragrance, mads (fruits containing large seeds surrounded by pulp (see Figure \ref{liste_fruits}), sold in the street mixed with sugar), bananas produced in small quantities and often imported from Ivory Coast, like pineapple.

\subsubsection{Drinks}
\begin{figure}[h!]
  \subfloat[bissap (hibiscus sabdariffa L.)]{
	\begin{minipage}[0.1\width]{0.35\textwidth}
	   \centering
	   \includegraphics[width=.8\textwidth]{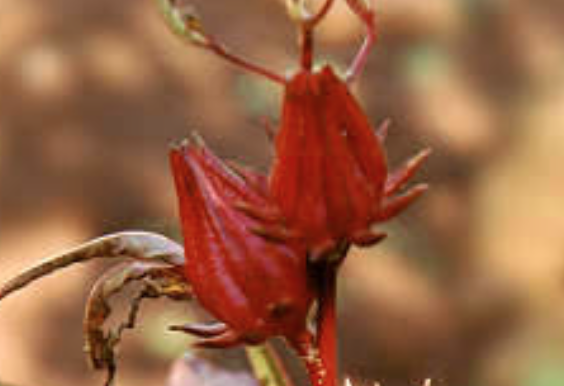}\label{bissap}
	\end{minipage}}
  \subfloat[bouye (Adansonia digitata L.)]{
	\begin{minipage}[0.1\width]{0.3\textwidth}
	   \centering
	   \includegraphics[width=.65\textwidth]{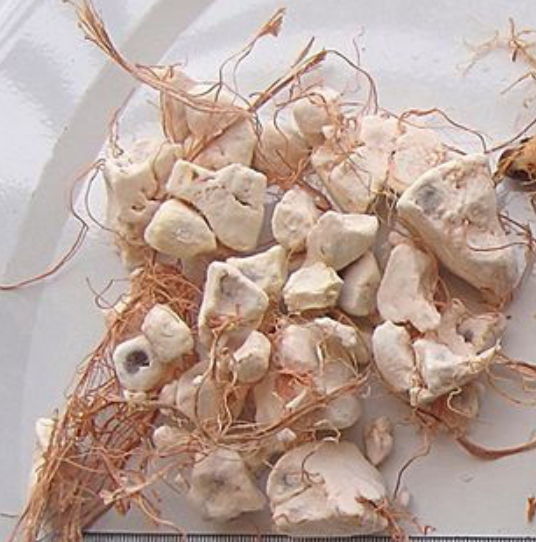}\label{bouye}
	\end{minipage}}
  \subfloat[mad (Saba senegalensis)]{
	\begin{minipage}[0.1\width]{ 0.35\textwidth}
	   \centering
	   \includegraphics[width=.7\textwidth]{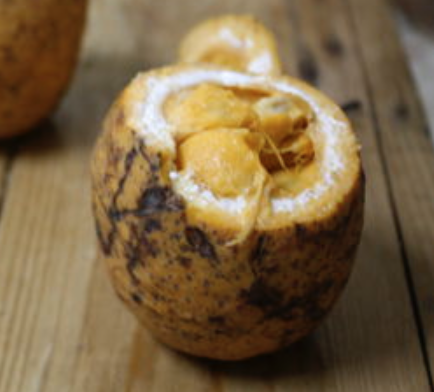}\label{mad}
	\end{minipage}}
\newline
   \subfloat[sump (Balanites aegyptiacus (L.) Delile)]{
	\begin{minipage}[0.1\width]{0.35\textwidth}
	   \centering
	   \includegraphics[width=.8\textwidth]{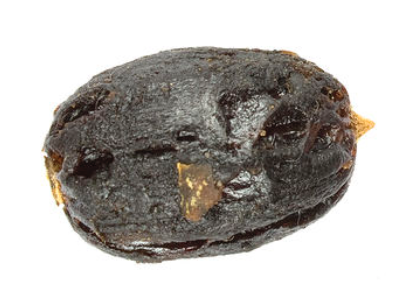}\label{sump}
	\end{minipage}}
  \subfloat[leng (Vitex doniana Sweet)]{  
	\begin{minipage}[0.1\width]{0.3\textwidth}
	   \centering
	   \includegraphics[width=.65\textwidth]{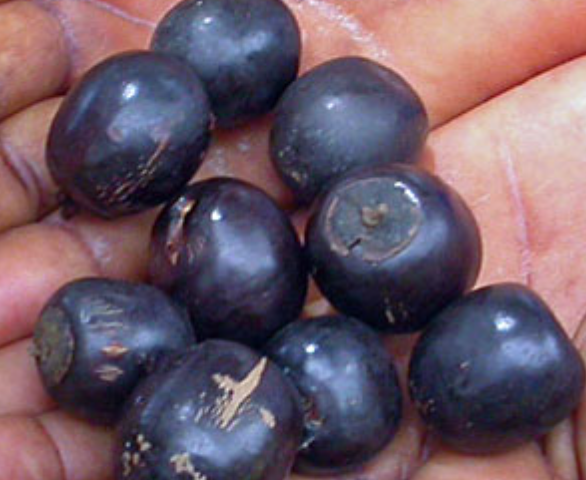}\label{leng}
	\end{minipage}}
  \subfloat[gerte tubaab (Terminalia catappa L.)]{
	\begin{minipage}[0.1\width]{ 0.35\textwidth}
	   \centering
	   \includegraphics[width=.7\textwidth]{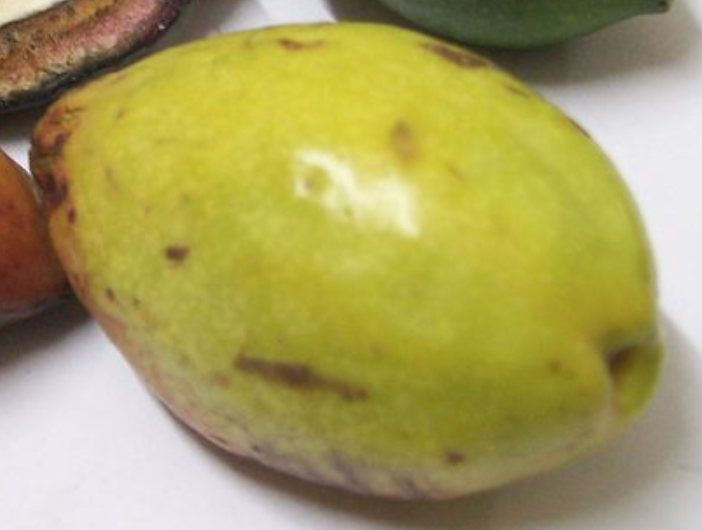}\label{gertetubaab}
	\end{minipage}}	
\caption{List of some fruits in Senegal (also used for drink) \cite{ausn}}\label{liste_fruits}
\end{figure}

\noindent The best known drink is undoubtedly bissap juice (hibiscus sabdariffa L.) decoction of red flower calyxes with a tangy taste) but you can also taste tamarind juice (dakkar) and ginger juice (ginger) also very widespread, the latter having a pungent taste that will not please everyone.\\
Street vendors and restaurants also offer drinks that are more original but more difficult to find: guava juice (buyap), mango, ditakh, sump, black plum,  etc. (see Figure \ref{liste_fruits}).\\
\noindent The pulp of the baobab fruit (bouye), called monkey bread (see Figure \ref{bouye}), a tree with multiple virtues, from leaves to seeds, is consumed as juice, an excellent rehydrating agent, and increasingly in the form of food supplements. Bouye is also used in the manufacture of cosmetics and diabetes medicines.\\ 
\noindent Plant contents oligo elements as zinc and selenium can modulate immunity with antioxidant properties. Polar extracts of many senegalese plants revealed potent antioxidant activity (\cite{sarr1, sarr2, Subhaswaraj}).

\subsection{Religions and beliefs of Senegal}
\noindent More than 90\% of the Senegalese population is of Muslim faith. The islamization of the country dates from the 11th century, when the Almoradives (Berber warrior monks) conquered northern Senegal. The appearance of Christianity is much more recent. Often mixed with its two religions, animism, with its rites and beliefs, is still very present.\\
Behind this low rate of contamination compared to the countries of Europe and the Americas, the simple religious ritual gestures could be among the salutes of the Senegalese people against the spread of the dreaded disease.\\ In New Castle (United Kingdom) a recent report published by Professor Richard Webber, an eminent academic, in collaboration with Trevor Philipps, a writer and former Labor politician, has come up with a very interesting observation: the ablutions of Muslims may have reduces the risk of contamination. The report comes as Public Health England launches an investigation into the reasons why non-whites seem to be the most affected by the disease (intensive care reports show that 34.5\% of critically ill patients are from ethnic minorities, although they only represent around 14\% of the population \cite{geog}. \\
\noindent In fact, if one of the keys to stopping the transmission of the virus is hand washing, a religious community in which all the faithful wash themselves thoroughly every day, and five times a day before performing the five daily prayers, by conforming to a very rigorous purification ritual, would it have anything to teach us? \\
\noindent Second, fasting is good for health and is said to boost our immunity (see \cite{Ayse, Mindikoglu}), and fasting twice a week is very good for your health. In addition to Ramadan, most Senegalese fast twice a week and on certain specific days of certain months of the Muslim calendar (Al-Hijira (Muharam 1), Lailat al Miraj (Rajab 27), Laylat Al Baraat (Sha'ban 15), Waqf Al Arafa - Hajj (Dhu'l-Hijjah 9). 

\subsection{Population density}\label{popdensity}
\begin{figure}[h!]
  \subfloat[Dakar north]{
	\begin{minipage}[1\width]{0.5\textwidth}
	   \centering
	   \includegraphics[width=1.\textwidth]{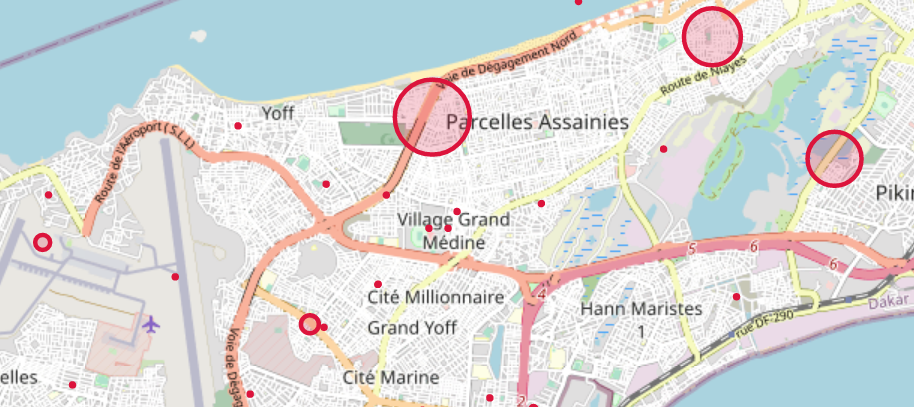}\label{dkrGPM}
	\end{minipage}}
  \subfloat[Dakar west and Dakar center]{
	\begin{minipage}[1\width]{ 0.5\textwidth}
	   \centering
	   \includegraphics[width=1.\textwidth]{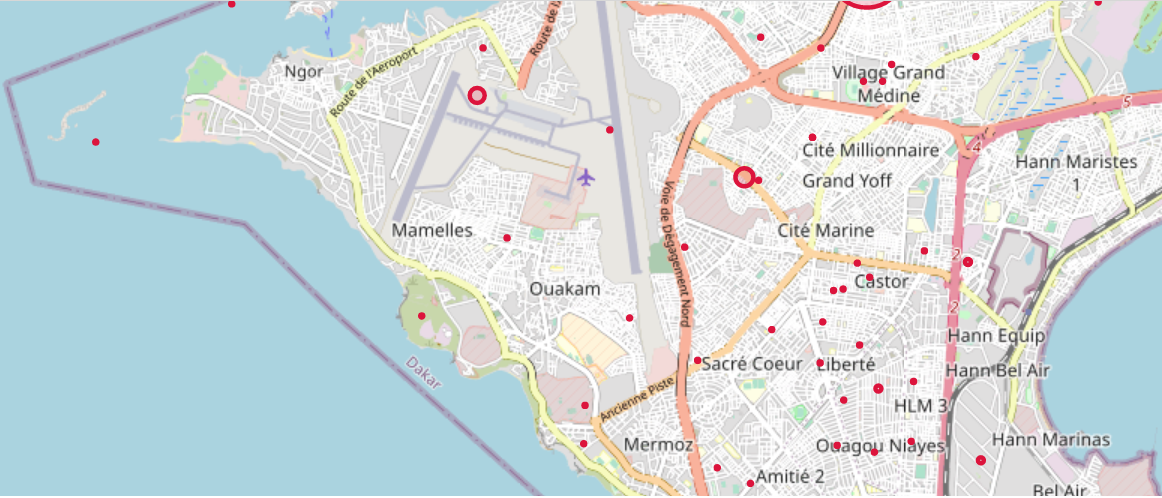}\label{dkr_center}
	\end{minipage}}
\newline
  \subfloat[Dakar south]{
	\begin{minipage}[1\width]{ 0.5\textwidth}
	   \centering
	   \includegraphics[width=1.\textwidth]{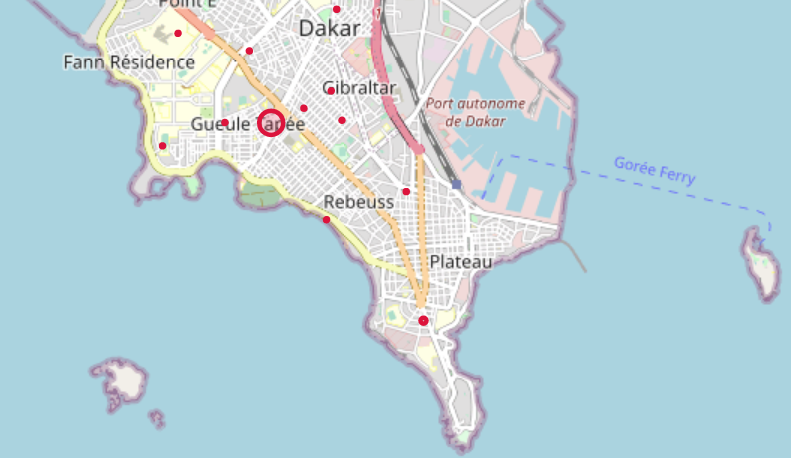}\label{dkr_sud}
	\end{minipage}}
%
  \subfloat[zoom of Dakar cases]{
	\begin{minipage}[1\width]{ 0.5\textwidth}
	   \centering
	   \includegraphics[width=1.\textwidth]{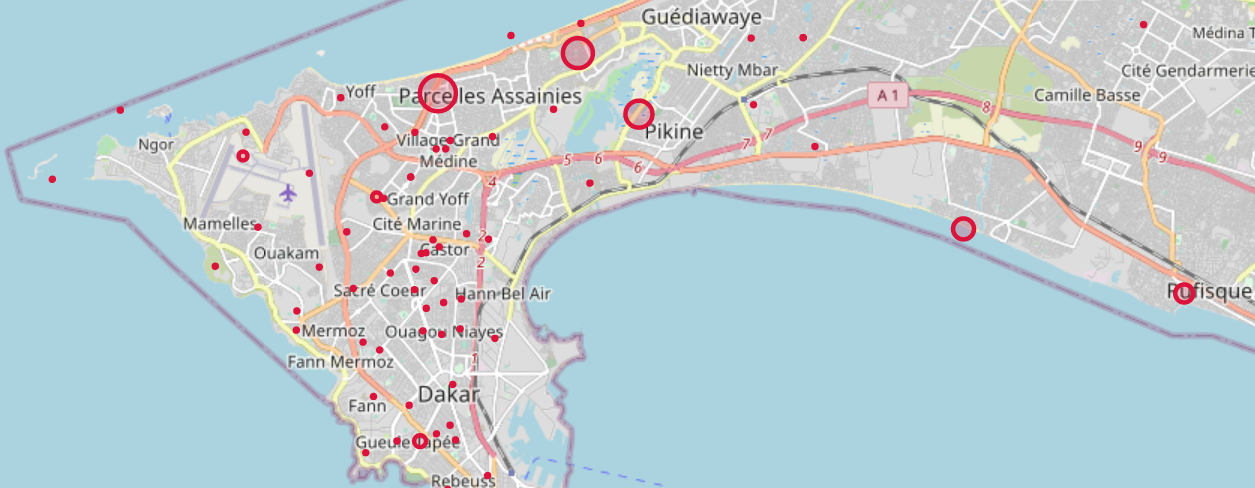}\label{dkr_zoom}
	\end{minipage}}
	
\caption{Dakar region cases}\label{dakar_viz}
\end{figure}

High population densities can catalyze the spread of COVID-19. With its 3,137,196 inhabitants, or almost a quarter of the population of Senegal (23.2\%), living on an area representing 0.3\% of the total area of the country, Dakar is the most populated region of Senegal and its population density is also the highest with 5,846 people/km2.\\
Keeping to more than 1-m distance between people coughing and sneezing, as recommended by the WHO \cite{who} becomes more difficult with higher population densities like in Dakar. Therefore, avoiding situations with higher population densities will be a necessary requirement to limit the spread of COVID-19 \cite{density}.\\
The relationship between the basic reproduction number, $R_0$, and the daily reproductive number, $\beta$, can be described by:
\begin{equation}\label{repronumber}
\beta = \tau c = \frac{R_0}{i}
\end{equation}
where $\tau$ denotes transmissibility and $c$ contact rate and where the infectious period $i$ equals one over the recovery rate $\gamma$.
This relationship holds for well-mixed populations, as assumed by standard compartmental models like SIR or SEIR which apply the law of mass action \cite{ndiaye1, ndiaye2, ndiaye3, balde}. They are also valid for small-to-medium spatial scales.\\
The contact rate is directly related to the total number of contacts, $C(t)$, generated for any person over a time period $t$, such that $C(t) = ctN$, for a situation with $N$ persons (3,137,196 for Dakar). This means that the effect of the time period of exposure and the crowd size to $C(t)$ are equivalent.\\
\noindent Parcelles Assainies, Guediewaye, etc. are zones with high density.  This also involves community cases (see Figure \ref{dkrGPM}). 
The urbanization rate is 44\%. By  June 12th, the infected cases in Dakar is 2727 (West (718), South  (684),  North (665) and Center (660)). This is not surprising, because the inhabitants of Dakar representing 23.5\% of Senegal population.  

\subsubsection{Simulations per zones}
\begin{figure}[h!]
  \subfloat[the first 5 zones]{
	\begin{minipage}[1\width]{0.5\textwidth}
	   \centering
	   \includegraphics[width=1.1\textwidth]{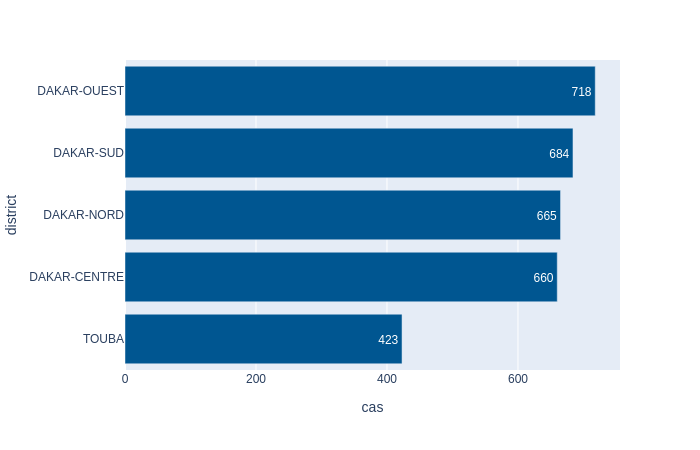}\label{first5_zones}
	\end{minipage}}
  \subfloat[the first 10 zones]{
	\begin{minipage}[1\width]{ 0.5\textwidth}
	   \centering
	   \includegraphics[width=1.1\textwidth]{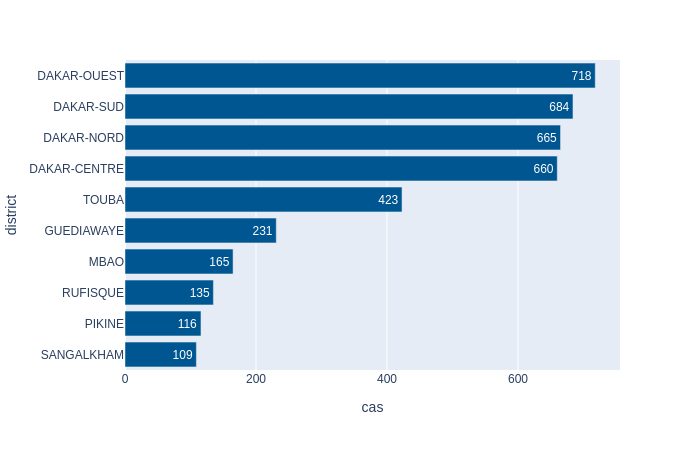}\label{first10_zones}
	\end{minipage}}
\caption{Number of confirmed cases in zones}\label{caseszones}
\end{figure}

\begin{figure}[h!]
	\centering
	\includegraphics[width=0.8\linewidth]{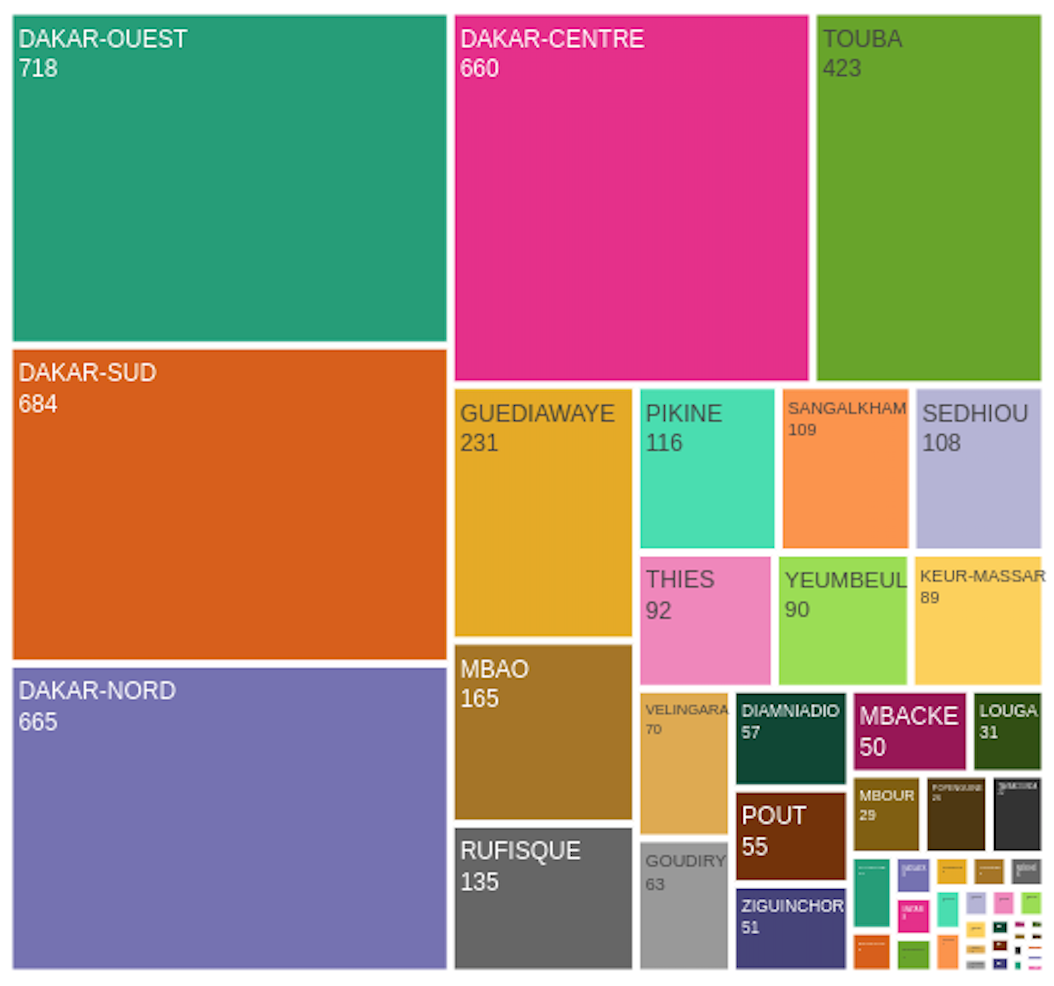}
	\caption{Cumulative number of confirmed cases per region}\label{cumsumregions}
\end{figure}

If we take the distribution of contamination cases, we see that Dakar (with high density area) is more contaminated. The figure \ref{dakar_viz} illustrates the visualization of high density areas (for the North, South, West and Center). 
Figure \ref{caseszones} illustrates this phenomenon (the first 5 (Figure \ref {first5_zones} and the first 10 (Figure \ref{first10_zones})). \\
\noindent In addition, the cumulative repartition of the number of confirmed cases per zone is given by Figure \ref{cumsumregions}.

\subsubsection{Simulations per district}
\begin{figure}[h!]
  \subfloat[the first 5 district]{
	\begin{minipage}[1\width]{0.5\textwidth}
	   \centering
	   \includegraphics[width=1.1\textwidth]{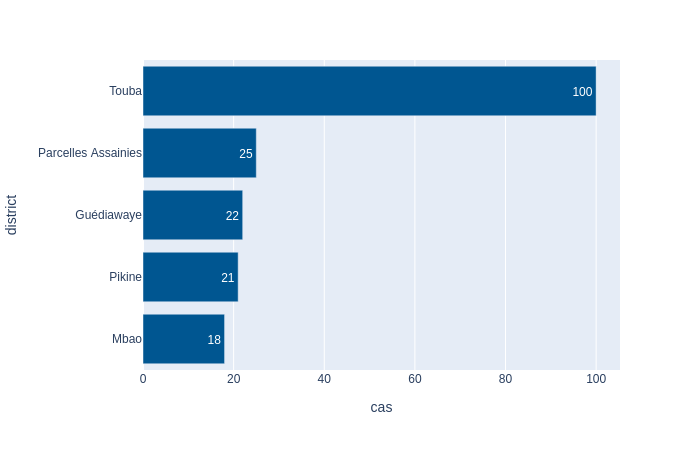}\label{first5_district}
	\end{minipage}}
  \subfloat[the first 10 district]{
	\begin{minipage}[1\width]{ 0.5\textwidth}
	   \centering
	   \includegraphics[width=1.1\textwidth]{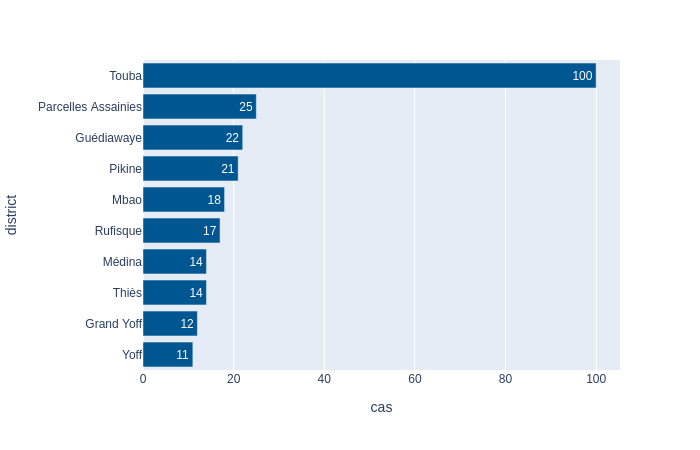}\label{first10_district}
	\end{minipage}}
\caption{Number of confirmed cases in districts}\label{casesdistricts}
\end{figure}

\begin{figure}[h!]
	\centering
	\includegraphics[width=0.8\linewidth]{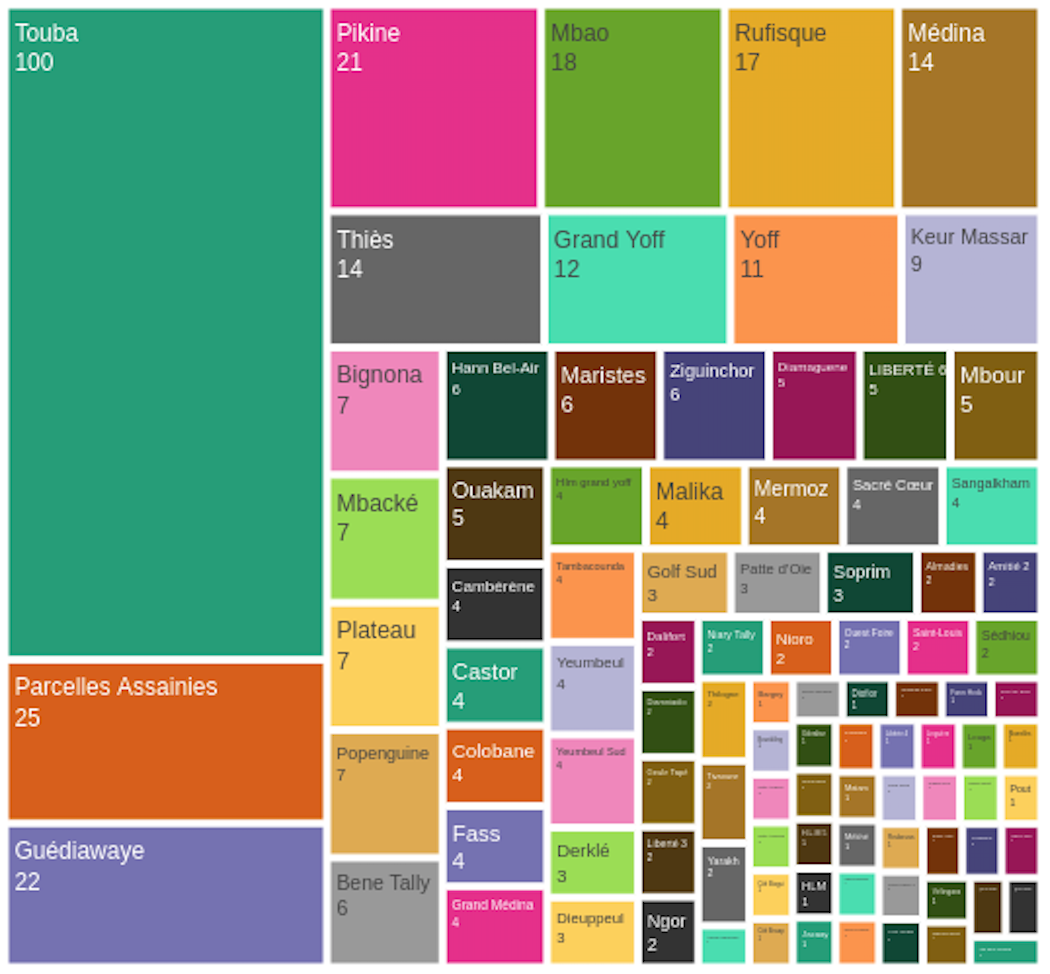}
	\caption{Cumulative number of confirmed cases per district}\label{cumsumdistrict}
\end{figure}

We perform the same simulation as in the previous section (by area). We see that Touba (in Figure \ref{regions_zoom1}), Parcelles Assainies, Guediewaye and Pikine (with high density area) are more contaminated. 
Figure \ref{casesdistricts} illustrates this phenomenon (the first 5 (Figure \ref {first5_district})  and the first 10 (Figure \ref{first10_district})). 
\noindent Also, the cumulative repartition of the number of confirmed cases per district is given by Figure \ref{cumsumdistrict}.

\section{Forecast (using Prophet) for the next 10 days}\label{simnum}
We perform one week ahead forecast with Prophet \cite{prophet}, with 95\% prediction intervals. Here, no tweaking of seasonality-related parameters and additional regressors are performed.
Prophet is a procedure for forecasting time series data based on an additive model where non-linear trends are fit with yearly, weekly, and daily seasonality, plus holiday effects. It works best with time series that have strong seasonal effects and several seasons of historical data. Prophet is robust to missing data and shifts in the trend, and typically handles outliers well. \\
\noindent For the average method, the forecasts of all future values are equal to the average (or “mean”) of the historical data. If we let the historical data be denoted by $y_1,...,y_T$, then we can write the forecasts as
$$
\hat{y}_{T+h|T}=\bar{y}=(y_1+y_2+...+y_T)/T
$$
The notation $\hat{y}_{T+h|T}$ is a short-hand for the estimate of $y_{T+h}$  based on the data $y_1,...,y_T$.\\
A prediction interval gives an interval within which we expect $y_t$  to lie with a specified probability. For example, assuming that the forecast errors are normally distributed, a 95\% prediction interval for the  $h$-step forecast is 
$
\hat{y}_{T+h|T}\pm1.96\hat{\sigma_h}
$
where  ${\sigma_h}$ is an estimate of the standard deviation of the $h$-step  forecast distribution.\\
\begin{table}
	\begin{center}
		\begin{tabular}{|c|c|c|c|} 
			\hline
			Date  &         $ \hat{y}$ &    $\hat{y}_{lower}$ &    $\hat{y}_{upper}$   \\
			\hline
 	2020-06-18 &	5344.432510& 	5265.770145 &	5416.670585\\
 	\hline
 	2020-06-19& 	5441.696552 &	5352.684245 &	5529.525833\\
 	\hline
 	2020-06-20 &	5532.082033 &	5422.459112 &	5632.964653\\
 	\hline
 	2020-06-21 &	5617.970335 &	5492.515978 &	5740.296607\\
 	\hline
 	2020-06-22& 	5715.813272 &	5576.778718 &	5848.876908\\
 	\hline
		\end{tabular}
	\end{center}
\caption{Senegal: predicted cumulative confirmed cases $\sim$June 22, 2020.}\label{sn_confcases}
\end{table} 
\begin{figure}[h!]
	\centering
	\includegraphics[width=0.8\linewidth]{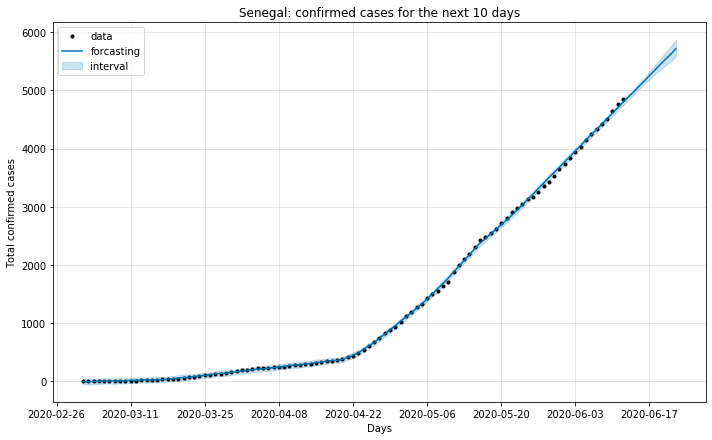}
	\caption{Senegal (Confirmed cases) : forecasting for the next 10 days $\sim$ June 22, 2020.}\label{sn_prediction}
\end{figure}
\noindent Most importantly, with the model and parameters in hand, we can carry out simulations for a longer time and forecast the potential trends of the COVID-19 pandemic. For  Figure \ref{sn_prediction}, the predicted cumulative number of confirmed cases is plotted for a shorter period of next 10 days.\\
\noindent We can summarize our basic predictions (from North to South), for all regions. At $\sim$June 22, 2020 we may obtain $>$  5700 confirmed cases (see Table \ref{sn_confcases}).\\
\noindent Finally, due to the inclusion of suspected cases with clinical diagnosis into confirmed cases (quarantined cases), we can see severe situation in some regions and/or districts, which requires much closer attention.

\section{Conclusion and perspectives}\label{ccl}
\begin{figure}[h!]
	\centering
	\includegraphics[width=0.5\linewidth]{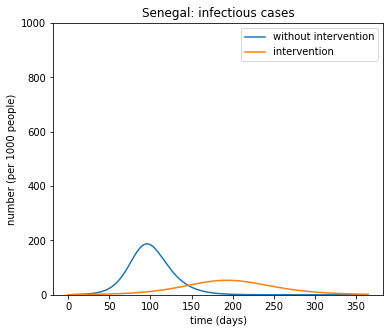}
	\vspace{-.25cm}
	\caption{The impact of infected cases with intervention}\label{intervent}
\end{figure}

Faced with the difficulty of enforcing social distancing measures, the systematic wearing of barrier masks and the multiple observations of favorable clinical evolution of confirmed cases for SARS-CoV2, it was logical and consistent to analyze and explore other contamination factors which are of interest in our tropical context.
This is how the most probable hypotheses arise: the youth of the population, the combined ambient temperature and humidity, acquired immunity (cross immunity induced by certain vaccines or similar previous infections, collective immunity, etc.).
Taking these factors into account should better model the course of the disease and not be limited to classical mathematical models for monitoring the course of epidemics.
An interdisciplinary approach seeks to validate new models inspired by the African context.
Based to our analysis, we can propose in future work, the generalized SIR  model \cite{ndiaye1, ndiaye2, ndiaye3, balde} taking into account to all these factors (cross immunity, youth, etc.) to analysis the coronavirus pandemic (COVID-19). \\ 
In addition, the authorities' decisions (social distancing, curfew, auto-discipline of senegalese people, etc.) may flat the curve (see Figure \ref{intervent}).

\subsection*{Acknowledgement}
The authors thanks the Non Linear Analysis, Geometry and Applications (NLAGA) project for supporting this work.


\begin{thebibliography}{10}
%
\bibitem{sarr1}
S.O. Sarr, A.D. Fall , R. Gueye, A. Diop, K. Diatta, N. Diop, B. Ndiaye,Y.M.  Diop, 2015. Etude de l’activité antioxydante des extraits de feuilles de Vitex doniana (Verbenaceae). International Journal of Biological and Chemical Science 2015; 9(3): 1263-1269.
%
\bibitem{sarr2}
S.O. Sarr, A.D. Fall , R. Gueye, A. Diop, B. Sene, K. Diatta, B. Ndiaye,Y.M.  Diop, 2015. Evaluation de l’activité antioxydante des extraits des feuilles de Aphania senegalensis (Sapindaceae) et de Saba senegalensis (Apocynaceae). 2015. International Journal of Biological and Chemical Science 2015; 9(6): 2676-2684.
%
\bibitem{Subhaswaraj} 
P. Subhaswaraj, M. Sowmya, V. Bhavana, M. Dyavaiah and B. Siddhardha. Determination of antioxidant activity of Hibiscus sabdariffa and Croton caudatus in Saccharomyces cerevisiae model system. J Food Sci Technol. 2017 Aug; 54(9): 2728–2736.
%
\bibitem{ndiaye1} 
B.M. Ndiaye, L. Tendeng, D. Seck, Analysis of the COVID-19 pandemic by SIR model and machine learning technics for forecasting, arXiv:2004.01574v1 [q-bio.PE], 3 Apr 2020, \url{https://arxiv.org/pdf/2004.01574.pdf}.
%
\bibitem{ndiaye2}  
B.M. Ndiaye, L. Tendeng, D. Seck, Comparative prediction of confirmed cases with COVID-19 pandemic by machine learning, deterministic and stochastic SIR models, arXiv:2004.13489 [q-bio.PE], 24 Apr 2020, \url{https://arxiv.org/pdf/2004.13489.pdf}. 
%
\bibitem{ndiaye3} 
M.A.M.T. Balde, C. Balde, B.M. Ndiaye, Impact studies of nationwide mea-
sures COVID-19 anti-pandemic: compartmental model and machine learning,
arXiv:2005.08395 [q-bio.PE], 17 May 2020. \url{https://arxiv.org/pdf/2005.08395.pdf}.
%
\bibitem{balde}
M.A.M.T. Bald\'e, Fitting SIR model to COVID-19 pandemic data and comparative forecasting with machine learning, medRxiv preprint doi: \url{https://doi.org/10.1101/2020.04.26.20081042. (2010)}.
%
\bibitem{ausn}
Au S\'en\'egal, Le coeur du S\'en\'egal, available on \url{https://www.au-senegal.com}. 
%
\bibitem{prophet} Prophet: Automatic Forecasting Procedure, avalailable in \url{https://facebook.github.io/prophet/docs/} or  \url{https://github.com/facebook/prophet}.
%
\bibitem{python} Python Software Foundation. Python Language Reference, version 2.7.  Available at \url{http} \\\url{://www.python.org}.
%
\bibitem{afri1} Coronavirus : Pourquoi l’Afrique est-elle globalement épargnée par l’épidémie ? available on: \url{https://www.20minutes.fr/monde/2778159-20200513-coronavirus-pourquoi-} \\ \url{afrique-globalement-epargnee-epidemie}.
%
\bibitem{msas} Minis\`ere de la sant\'e et de l'action sociale, May 31, 2020, available on \url{http://www.sante.} \\\url{gouv.sn/}.
%
\bibitem{pluie1} Variabilité pluviométrique au sénégal, 2020, available on \url{https://www.google.com/} \\\url{search?q=variabilit\%C3\%A9+pluviom\%C3\%A9trique+au+s\%C3\%A9n\%C3\%A9gal&sxsrf=ALeKk00rQ2UyTNzgCzlzf-NtS2yLSDS6rQ:1590842365018&source=lnms&tbm=isch&sa=X&ved=2ahUKEwiQzb3QzdvpAhWQsBQKHauqCmMQ_AUoAnoECAwQBA&biw=1366&bih=625#imgrc=jZnLBB_QuKfg6M&imgdii=yTNEQZvKam591M}.
%
\bibitem{immunity1} Ray M-C. Immunité croisée, Futura, May 30, 2020, available on \url{https://www.futura-sciences.com/sante/definitions/medecine-immunite-croisee-16572/}.
%
\bibitem{immunity2} Pouvez-vous attraper le coronavirus deux fois? FR24 News France. May 2020, available on \url{https://www.fr24news.com/fr/a/2020/05/commentaire-pouvez-vous-attraper-} \\\url{le-coronavirus-deux-fois.html}.
%
\bibitem{immunity3} Coronavirus : d\'ecryptage des hypothèses qui expliqueraient la faible contamination en Afrique - Jeune Afrique, 2020, available on \url{https://www.jeuneafrique.com/937712/societe/coronavirus-decryptage-des-hypotheses-qui-expliqueraient-la-} \\\url{faible-contamination-en-afrique/}.
%
\bibitem{population2} 52,1\% de la population du Sénégal a moins de 20 ans,  available on \url{ http://www.ansd.sn/} \\\url{index.php?option=com_content&view=article&id=608:2020-05-14-10-04-38&ca} \\\url{tid=56:depeches&Itemid=264}.
%
\bibitem{geog} Données géographiques et identité religieuse au Sénégal. La Croix Africa, available on \url{https://africa.la-croix.com/statistiques/senegal/}.
%
\bibitem{density}
Joacim Rocklov, Henrik Sjodin, High population densities catalyse the spread of COVID-19, Journal of Travel Medicine, 29 March 2020, 1–2 doi: 10.1093/jtm/taaa038. 
%
\bibitem{who} WHO, Coronavirus disease(COVID-19) advice for the public. \url{https:// www.Who.Int/emergencies/diseases/novel- coronavirus- 2019/advice- for-public}.
%
\bibitem{R-value} Beihang University COVID-19 Research Team, available on  \url{http://covid19-report.com/#/r-value}

\bibitem{Ayse}
A.L.Mindikoglu, M.M. Abdulsada, A. Jain, J. Min Choi, P.K. Jalal Sridevi Devaraj, M.P. Mezzari, J.F. Petrosino, A.R. Opekun, S. Yun Jung. 
Intermittent Fasting From Dawn to Sunset for 30 Consecutive Days Is Associated With Anticancer Proteomic Signature and Upregulates Key Regulatory Proteins of Glucose and Lipid Metabolism, Circadian Clock, DNA Repair, Cytoskeleton Remodeling, Immune System and Cognitive Function in Healthy Subjects, Journal of Proteomics, 2020 Apr 15, 217:103645.  doi: 10.1016/j.jprot.2020.103645.

\bibitem{Mindikoglu}
A.L. Mindikoglu, Impact of time-restricted feeding and dawn-to-sunset fasting on circadian rhythm, obesity, metabolic syndrome, and nonalcoholic fatty liver disease, Gastroenterol. Res. Pract. 2017 (2017) 3932491.

\end{thebibliography}
\end{document}